\documentclass{PoS}

\title{On description of the correlation between multiplicities in windows separated\\ in azimuth and rapidity}

\ShortTitle{On description of the correlation between multiplicities}

\author{\speaker{Vladimir Vechernin}%
         \thanks{The work was supported by the RFFI grant 12-02-00356-a.}\\
        Saint-Petersburg State University\\
        E-mail: \email{vechernin@pobox.spbu.ru}}
\abstract{
The forward-backward (FB) multiplicity correlations in two windows  separated in rapidity and azimuth are analyzed in the framework of the model with independent identical emitters (strings). Along with the short-range contribution, originating from the correlation between particles produced by a single string, the long-range contribution, originating from the fluctuation in the number of strings, is taken into account. The connection of the FB multiplicity correlation coefficient with the two-particle correlation function and the di-hadron correlation analysis is traced. It's also shown that the direct azimuthal flow leads to the forward ridge structure in the resulting two-particle correlation function.
}

\FullConference{The XXI International Workshop High Energy Physics and Quantum Field Theory,\\
		June 23 -- June 30, 2013 \\
		Saint Petersburg Area, Russia}
\def\nF{{n_F^{}}}

\def\nB{{n_B^{}}}

\def\nF{{n_F^{}}}
\def\nB{{n_B^{}}}

\def\mo{{\mu^{}_{0}}}

\def\rhoo{{\rho^{}_{0}}}

\def\yF{{\eta^{}_F}}
\def\yB{{\eta^{}_B}}

\def\fv{\phi}

\def\loyo{\lambda_1(\eta_1)}
\def\loyt{\lambda_1(\eta_2)}

\def\lo#1{\lambda_1(#1)}

\def\lt#1{\lambda_2(#1)}
\def\Lam#1{\Lambda(#1)}

\def\loyf{\lambda_1(\eta,\phi)}

\def\ltyf{\lambda_2(\eta_1,\phi_1;\eta_2,\phi_2)}

\def\ro#1{\rho_1(#1)}

\def\bc{\begin{center}}
\def\ec{\end{center}}
\def\beq{\begin{equation}}
\def\eeq{\end{equation}}

\def\hs#1{\hspace*{#1cm}}
\def\av#1{\langle #1 \rangle}

\def\tg{{\textrm{tg}}\,}

\def\dyF{{\delta\eta^{}_F}}
\def\dyB{{\delta\eta^{}_B}}

\def\dfF{{\delta\phi^{}_F}}
\def\dfB{{\delta\phi^{}_B}}

\begin{document}

%%%%%%%%%%%%%%%%%%%%%%%%%%%%%%%%%%%%%%%
%%%%%%%%%%%%%%%%%%%%%%%%%%%%%%%%%%%%%%%
%%%%%%%%%%%%%%%%%%%%%%%%%%%%%%%%%%%%%%%
\section{Connection of the FB correlation coefficient with two-particle correlation function}  \label{bC2}
Usually under the forward-backward (FB) correlation one implies the correlation between
the multiplicities of charged particles $\nF$ and $\nB$ in two separated rapidity windows
$\dyF$ and $\dyB$ in high energy pp, pA or AA interactions. In present report we consider
a more general case - the FB correlation in windows separated both in rapidity and in azimuth,
when two azimuthal sectors $\dfF$ and $\dfB$ are selected within these FB windows $\dyF$ and $\dyB$.

Traditionally one uses the following definition of the FB correlation coefficient \cite{PPR}:
\begin{equation} \label{b}
b_{abs}\equiv\frac{\av{\nF\nB}-\av{\nF}\av{\nB}}{\av{\nF^2}-\av{\nF}^2}
 \hs1 \textrm{or}
 \hs1 b_{rel}\equiv\frac{\av{\nF}}{\av{\nB}}b_{abs}
\end{equation}
The last one is using, when the analysis is performed in so-called relative or scaled variables \cite{PoS12Feof},
i.e. for the correlation between the normalized values $\nF/\av{\nF}$ and $\nB/\av{\nB}$.

The two-particle correlation function $C_2$ is defined through the inclusive $\rho_1$
and double inclusive $\rho_2$ distributions \cite{Voloshin02}. If we consider the distributions,
integrated over the absolute value of transverse momenta,  we have
\begin{equation}
\label{C2}
C_2(\eta_F, \phi_F;\eta_B, \phi_B)\equiv\frac{\rho_2 (\eta_F, \phi_F;\eta_B, \phi_B)}{\rho_1(\eta_F, \phi_F) \rho_1 (\eta_B, \phi_B)}-1
\end{equation}
\begin{equation}
\label{rho21}
\rho_1 (\eta, \phi)=\frac{d^2N}{d\eta\,d\phi}
\ , \hs1
\rho_2 (\eta_F, \phi_F;\eta_B, \phi_B)=\frac{d^4N}{d\eta_F\,d\phi_F\,d\eta_B\,d\phi_B}
\end{equation}
In experiment one measures the  $\rho_1$ taking a small window
$\delta\eta\,\delta\phi$ around $\eta$, $\phi$, then
\begin{equation}
\label{rhoex1}
\rho_1(\eta,\phi)=\frac{\av{n}}{\delta\eta\,\delta\phi}  \ ,
\end{equation}
here $\av{n}$ is the mean multiplicity in the acceptance $\delta\eta\,\delta\phi$.
%One has to reduce the acceptance until the ratio (\ref{rhoex1}) becomes constant.
Similarly, by definition (\ref{rho21}) to measure the  $\rho_2$  one has  to take  two small windows:
$\delta\eta_F\,\delta\phi_F$ around $\eta_F$, $\phi_F$  and
$\delta\eta_B\,\delta\phi_B$ around $\eta_B$, $\phi_B$, then
\begin{equation}
\label{rhoex2}
\rho_2(\eta_F,\phi_F;\eta_B,\phi_B) =\frac{\av{\nF\nB}}{\delta\eta_F\,\delta\phi_F\,\delta\eta_B\,\delta\phi_B} \ .
\end{equation}
%One has to reduce the acceptances of the observation windows until the ratio (\ref{rhoex2}) becomes constant.
The formulae (\ref{rhoex1}) and (\ref{rhoex2}) are the base for the experimental measurement of
the one- and two-particle densities of charge particles. By (\ref{rhoex1}) and (\ref{rhoex2}) the definition (\ref{C2})  leads
to the following experimental procedure of the determination of the correlation function $C_2$:
\begin{equation} \label{C2ex}
C_2(\eta_F, \phi_F;\eta_B, \phi_B)=\frac{\av{\nF\nB}-\av{\nF}\av{\nB}}{\av{\nF}\av{\nB}}
%= \avL{\nFr\nBr}-1
 \ ,
\end{equation}
where $\nF$ and $\nB$ are the event multiplicities in two small windows: $\delta\eta_F\,\delta\phi_F$ and $\delta\eta_B\,\delta\phi_B$.

%%%%%%%%%%%%%%%%%%%%%%%%%%%%%%%%%%%%%%%
Comparing (\ref{b}) and (\ref{C2ex}) we see that for small FB windows we have
\begin{equation} \label{b-C2}
b_{abs}=\frac{\av{\nF}\av{\nB}}{D_\nF} C_2(\eta_F, \phi_F;\eta_B, \phi_B)
 \ , \hs1
b_{rel}=\frac{\av{\nF}^2}{D_\nF} C_2(\eta_F, \phi_F;\eta_B, \phi_B) ,
\end{equation}
where $D_\nF=\av{\nF^2}-\av{\nF}^2$. Note that for small forward window: $D_\nF\to\av{\nF}$ \cite{FB13}.
So we see that the traditional definition (\ref{b}) of the FB correlation coefficient
in the case of two small observation windows, separated in azimuth and rapidity,
coincides with the standard definition of two-particle correlation function $C_2$
upto some common factor $\av{\nB}$ or $\av{\nF}$, which depends on the width of windows.

 %%%\section{Untriggered  di-hadron correlation} \label{dihadr} %

In practice, in di-hadron correlation analysis, the following
alternative definition of the two-particle correlation function $C$ is in use
\cite{STARdihadr09,CMSridge10}:
\beq \label{C-SB} \label{S}
C=\frac{S}{B}-1 \ , \hs1
S=\frac{d^2 N}{d\Delta \eta\, d\Delta\phi}    \ ,
\eeq
where $\Delta \eta=\eta_1-\eta_2$ and  $\Delta\phi=\phi_1-\phi_2$
are the distances between two particles in rapidity and in azimuth,
and one takes into account all possible pair combinations of particles produced
in given event in some one large rapidity interval $\Delta\eta\in(Y_1,Y_2)$.
The $B$  is the same, but in the case of uncorrelated particle production,
obtained by the event mixing procedure.

At such definition, in contrast with (\ref{C2}), one implies from the very beginning
that the translation invariance in rapidity takes place and the result depends
only on $\Delta \eta=\eta_1-\eta_2$ for any $\eta_1,\eta_2 \in (Y_1,Y_2)$.
(All the pairs with the same value of difference $\eta_1-\eta_2$
contribute to the same bin of the multiplicity distribution, irrespective of
%regardless of
the value of  $(\eta_1+\eta_2)/2$.)
This assumption is reasonable only in the central rapidity region at high energies.
It means that we suppose that  in the interval $(Y_1,Y_2)$:
\beq \label{ro-trans}
\ro{\eta}=\rhoo    \ , \hs1
 \rho^{}_2(\eta_1,\eta_2;\Delta\phi)= \rho^{}_2(\eta_1-\eta_2,\Delta\phi)   \ .
\eeq
In this case we have  for the enumerator of (\ref{C-SB}):
\beq \label{S1}
S(\Delta \eta,\Delta\phi)=  \int_{Y_1}^{Y_2} dy_1\, dy_2\, \rho^{}_2(y_1-y_2,\Delta\phi)\, \delta(y_1-y_2-\Delta \eta)    \ ,
\eeq
or in the case of commonly used symmetric interval $(-Y/2,Y/2)$:
\beq \label{S2}
S(\Delta \eta,\Delta\phi)= \rho^{}_2(\Delta \eta,\Delta\phi)\,  t_Y^{} (\Delta \eta)
\eeq
where  the  $t_Y^{}(\Delta \eta)$ is  the "triangular" weight function:
\beq \label{tri}
t_Y^{} (\Delta \eta) = [\theta(-\Delta \eta)(Y+\Delta \eta) +\theta(\Delta \eta)(Y-\Delta \eta)]\,\theta(Y-|\Delta \eta|) \ .
\eeq

In the denominator of (\ref{C-SB}) for mixed events we should replace the $\rho^{}_2(\eta_1,\eta_2,\Delta\phi)$
by the product $\rho^{}_1(\eta_1) \rho^{}_1(\eta_2)$,
which due to the translation invariance in rapidity reduces simply to $\rho^{2}_0$.
Then
\beq \label{B2}
B(\Delta \eta,\Delta\phi) = \rho^{2}_0 \   t_Y^{} (\Delta \eta) \ .
\eeq
Substituting into (\ref{C-SB}) we get
\beq \label{C-C2}   \label{RC2a}
C(\Delta \eta,\Delta\phi)=   \frac{\rho^{}_2(\Delta \eta,\Delta\phi)}{\rho^{2}_0}-1=C^{}_2(\Delta \eta,\Delta\phi) \ ,
\eeq
We see if the translation invariance in rapidity takes place within the interval $(Y_1,Y_2)$,
then the definition (\ref{C-SB}) is equivalent to the standard one  (\ref{C2})
(see meanwhile the remark in the end of the next section).

The drawback of this approach is that it supposes from the very beginning the translation invariance
and hence can't be applied for an investigation of the multiplicity correlation at
large rapidity distances, where the translation invariance in rapidity  (\ref{ro-trans}) is not valid.
At that by (\ref{C2ex}) and (\ref{b-C2}) we see
that the approaches based on the analysis of the standard (\ref{b})
FB correlation coefficient
with two remote windows of small acceptance in rapidity and azimuth
enable in any case to measure the  correlation function  $C_2(\eta_1,\eta_2;\phi_1-\phi_2)$
without using of the event mixing procedure.

%%%%%%%%%%%%%%%%%%%%%%%%%%%%%%%%%%%%%%%
\section{Model with strings as independent identical emitters} \label{strings}
%%%%%%%%%%%%%%%%%%%%%%%%%%%%%%%%%%%%%%%
We now calculate the FB correlations in windows separated
in rapidity and azimuth using the simple two stage model \cite{PLB00,EPJC04,YF1},
inspired by a string picture of hadronic interactions.
In this model one suggests that at the initial stage of interaction some number $N$ of strings are formed,
which fluctuates event-by-event with some variance $D_N=\av {N^2}-\av {N}^2$ or scaled variance
\beq \label{omN}
\omega_N=D_N/\av N  \ .
\eeq
Note that the fluctuation in the number of strings
in pp and especially in AA collisions is not poissonian  \cite{PRC11}
and hence $\omega_N\neq 1$. Its value depends on the collision energy.
At next stage one considers these strings as identical independent emitters of observed charge particles.

In the present paper, along
with the so-called long-range (LR) part of the correlation \cite{CapKrz},
originating from the fluctuation in the number of strings,
we take into account  also
the short-range (SR) contribution, originating from the correlation between
particles produced by a single string.

To characterize the last property of the string we introduce,
similarly to the consideration in the section \ref{bC2},
the $\loyf$ and $\ltyf$ - the one- and two-particle densities of charge particles produced by one string.
In this section we'll suppose that the particle emission from one string is isotropic in $\phi$,
then at fixed number of strings ($N$) in the framework of the model we have:
\beq \label{roNo}
\rho^{N}_1(\eta)=N  \lambda_1(\eta) \ ,
\eeq
\beq \label{roNt}
\rho^{N}_2(\yF ,\yB ;\Delta\phi)=N \lambda_2(\yF ,\yB ;\Delta\phi) + N(N-1)\lambda_1(\yF ) \lambda_1(\yB )   \ .
\eeq
After averaging over $N$ the one- and two-particle densities of charge particles  are given by
\beq \label{ro1-av}
\rho^{}_1(\eta)=\av{N}\lambda_1(\eta) \ ,
\eeq
\beq \label{ro2-av}
\rho^{}_2(\yF ,\yB ;\Delta\phi)=
\av N [\lambda_2(\yF ,\yB ;\Delta\phi) -\lambda_1(\yF ) \lambda_1(\yB )]+ \av{N_{}^2}\lambda_1(\yF ) \lambda_1(\yB )
\ . \eeq
%and
%\beq \label{cor-av}
%\rho^{}_2(\yF ,\yB ;\Delta\phi)-\rho^{}_1(\yF )\rho^{}_1(\yB )=
%%%%%%%%%\av{ \rho^{}_2(\yF ,\yB ;\Delta\phi) }-\av{ \rho^{}_1(\yF ) }\av{ \rho^{}_1(\yB ) }=
%\eeq
%$$
%=\av N [(\lambda_2(\yF ,\yB ;\Delta\phi) -\lambda_1(\yF ) \lambda_1(\yB )]+ {  D_N^{}} \lambda_1(\yF ) \lambda_1(\yB )   \ .
%$$
%%%%%%%%where $D_{N}$ is {  the event-by-event variance $D_N=\av{N^2}-\av{N}^2 $ of the number of emitters}.
%%%%%%%%%%%%%%%%%%%%%%%%%%%%%%%%%%%%%%%
%%%%%%%%where $\omega_N$ is {the event-by-event scaled variance $\omega_N=D_N/\av N $ of the number of emitters}
Introducing  similarly (\ref{C2}) the two-particle correlation function for charged particles
produced from a decay of a single string:
\beq \label{Lam1}
\Lambda(\eta_1,\eta_2;\Delta\phi)=\frac{\lt{\eta_1,\eta_2;\Delta\phi}}{\loyo\loyt}-1 \ ,
\eeq
we find for the two-particle correlation function $C_2$ the following expression:
\begin{equation}  \label{C2Lam}
C_2(\yF ,\yB ;\Delta\phi)=\frac{\Lam{\yF ,\yB ;\Delta\phi} +{  \omega_N}  }{\av N} \ ,
\end{equation}

%By (\ref{D_mF}) we see that the presence of SR correlation
%turns the string into non-poissonian emitter.

In the central rapidity region, where one has the translation invariance in rapidity
and each string contributes to the particle production in the whole rapidity region,
we have
\beq \label{Lam2}
\lo{\eta}=\mo=const    \ , \hs1
\Lam{\yF ,\yB ;\Delta\phi}=\Lam{\yF -\yB ,\Delta\phi}   \ ,
\eeq
then
\beq \label{ro1-av1}
\rho^{}_1(\eta)=\av{N}\mo=const \ ,
\eeq
\begin{equation}  \label{C2Lam1}        \label{C-C2a}
C_2(\Delta\eta, \Delta\phi)=\frac{\Lambda(\Delta\eta, \Delta\phi)+ \omega_N}{\av N} \ .
\end{equation}
%Recall that $\Delta\eta$ and $\Delta\phi$
%are the distances between the centers of forward and backward windows in rapidity and azimuth.
Important that by (\ref{C2Lam1}) we see  that the value of the common constant (pedestal)
in $C_2(\Delta\eta, \Delta\phi)$ is physically important.
The height of the pedestal ($\omega_N/\av N$=$D_N/\av {N}^2$)
contains the important physical information
on the magnitude of the fluctuation of the number of emitters N at different energies and centrality fixation
\cite{FB13,CapKrz}.

%===%%%%%%%%%%%%%%%%%%%%%%%%%%%%%%%%%%%%%%%
In a conclusion of the section we note that if one uses the so-called di-hadron correlation approach,
described above, for the experimental determination
of the two-particle correlation function  $C(\Delta \eta,\Delta\fv)$   (\ref{C-SB})
the result can depend on the details of track and/or event mixing used in that approach
for the determination of $B$ through the imitation of the uncorrelated particle production
and also on arbitrary using of unjustified normalization procedure in $S$ and $B$.

One can illustrate this in the framework of the model with strings as independent identical emitters.
By (\ref{S2}) and (\ref{ro2-av}) we have for the enumerator and the denominator of (\ref{C-SB}):
\beq \label{ro2avxx}
S(\Delta \eta,\Delta\phi)
 =\av{\rho^{N}_2(\Delta \eta;\Delta\phi)}\,  t_Y^{} (\Delta \eta)
=[\av N \Lam{\Delta \eta,\Delta\phi}+ \av{N_{}^2}]\mu_0^{2}\,  t_Y^{} (\Delta \eta)   \ ,
\eeq
\beq \label{B2a}
B(\Delta \eta,\Delta\phi)
=\int_{-Y/2}^{Y/2} dy_1\, dy_2\, \av{\rho^{N}_1(y_1) }\av{ \rho^{N}_1(y_2)}\, \delta(y_1-y_2-\Delta \eta)
= \av N^2 \mu^{2}_0 \  t_Y^{} (\Delta \eta) \ .
\eeq
Then by $C=S/B-1$ we get again that $C(\Delta \eta,\Delta\phi)=C_2(\Delta \eta,\Delta\phi)$, which is given by (\ref{C2Lam1}).
%\beq \label{C-C2a}
%C(\Delta \eta,\Delta\phi)=\frac{\omega_N+ \Lam{\Delta \eta,\Delta\phi}}{\av N}=C_2(\Delta \eta,\Delta\phi) \ .
%\eeq

But if instead of (\ref{B2a}) one will use another event mixing procedure, for example, the mixing
only between events with the same multiplicity
(i.e. the same $N$), then instead of  (\ref{B2a}) we'll have
\beq \label{B2b}
B(\Delta \eta,\Delta\fv) =\int_{-Y/2}^{Y/2} dy_1\, dy_2\, \av{\rho^{N}_1(y_1)  \rho^{N}_1(y_2)}\, \delta(y_1-y_2-\Delta \eta)
= \av {N^2} \mu^{2}_0 \  t_Y^{} (\Delta \eta) \ ,
\eeq
which leads instead of  (\ref{C-C2a}) to
\beq \label{C-C2b}
C(\Delta \eta,\Delta\fv)=\frac{\av N}{\av {N^2}}\Lam{\Delta \eta,\Delta\fv} \ .
\eeq

The last result does not coincide
with the standard two-particle correlation function $C_2$, defined by (\ref{C2}).
Compare (\ref{C-C2b}) with (\ref{C-C2a}) we see that in this case the resulting $C(\Delta \eta,\Delta\fv)$
does not have an additional constant contribution reflecting  the event-by-event fluctuation in the number of emitters.
It depends only on the pair correlation function of a single string $\Lam{\Delta \eta,\Delta\fv}$ and, therefore,
is equal to zero in the absence of the
pair correlation from one string.

The same effect can take place if one use unjustified artificial normalization procedure in $S$ and $B$.

%%%%%%%%%%%%%%%%%%%%%%%%%%%%%%%%%%%%%%%
%\begin{figure}
%\centering
%\pause
%\includegraphics[width=0.8\linewidth]{SFM.eps}
%\vspace{0.2cm}
%\end{figure}

%%%%%%%%%%%%%%%%%%%%%%%%%%%%%%%%%%%%%%%
\section{Connection between the ridge and the azimuthal flows}

In this section we consider a simple model, in which we will not take into account the
two-particle correlation between particles originating from
the decay of a same string ($\Lambda(\Delta\eta, \Delta\phi)=0$),
but try to understand the influence of the event-by-event fluctuation
of azimuthal distribution on the resulting two-particle correlation function.
The physical reason which leads to the event anisotropy of the azimuthal distribution,
for example in the framework of the string fusion approach \cite{SF},
is the final state interaction (FSI) of produced particles with the fused string medium.
%the fluctuation of string distribution in the transverse plane, of string configuration
%the influence the possible effects of the interaction between strings, e.g. in the form of their fusion

In papers \cite{v2,flows} in the framework of this approach the azimuthal flows $v_n$
for ultrarelativistic heavy ion collisions were found
by Monte Carlo (MC) simulations.
In that papers the flows were calculated by
Fourier decomposition of the azimuthal inclusive distribution of charged particles $\rho^{i}_1(\phi)$, produced
by the given string configuration $i$, obtaining by MC simulations
(it was supposed that in the central region this configurations are homogeneous in rapidity):
\beq
\label{rhoi}
\rho_1^i(\phi)=\bar{\rho}^i[1+2\sum_{n=1}^\infty (a^i_n \cos n\phi + b^i_n \sin n\phi)]
=\bar{\rho}^i [1+2\sum_{n=1}^\infty v^i_n \cos n(\phi - \psi^i_n)]   \  ,
\eeq
where
\beq
\label{ain}
a^i_n=\frac{1}{2\pi\bar{\rho}^i}\int \rho_1^i(\phi) \cos n\phi \, d\phi \ ,  \hs1
\label{bin}
b^i_n=\frac{1}{2\pi\bar{\rho}^i}\int \rho_1^i(\phi) \sin n\phi \, d\phi  \ ,
\eeq
\beq
\label{vin}
\bar{\rho}^i=\frac{1}{2\pi}\int   \rho_1^i(\phi) d\phi \ ,   \hs1
\label{rhoi0}
v^i_n=\sqrt{a^{i2}_n +b^{i2}_n} \ , \hs1 \tg n\psi^i_n=b^i_n/a^i_n     \ ,
\eeq
Here $\bar{\rho}^i$ is the mean multiplicity for given string configuration.
The flows were founded by averaging over string configurations $i=1,...,K$:
\beq
\label{vn}
v_n=\frac{1}{K}\sum_{i=1}^K  v^i_n
=\frac{1}{K}\sum_{i=1}^K  \sqrt{a^{i2}_n +b^{i2}_n} \ .
\eeq

Since there are no correlations between particles produced
by different strings and
we don't  take into account the
correlations between particles originating from
the decay of a same string,
then for a given string configuration $i$, we have
\beq \label{factor}
\rho^{i}_2(\phi_1,\phi_2)=\rho^{i}_1(\phi_1)\rho^{i}_1(\phi_2)  \ .
\eeq
We'll show now that nevertheless one has in this model
the so-called ridge structure in the two-particle correlation function $C_2$,
which can be expressed  through the same Fourier harmonics $a_n^i$ and $b_n^i$,
as the azimuthal flows $v_n$.

As discussed, in the central rapidity region di-hadron correlation function is given by the expression (\ref{RC2a}).
%(see, for example, V.V. Vechernin, Arxiv:1305.0857).
In the framework of this model we have for the di-hadron correlation function:
\beq \label{RC2}
C(\Delta \phi)=C_2(\phi_1-\phi_2)=\frac{\rho_2(\phi_1-\phi_2)} {\rho^2_1} -1  \ ,
\eeq
where  $\rho_1$ is the mean multiplicity density:
\beq \label{rho1a} \label{rho1}
\rho_1
= \frac{1}{K}\sum_{i=1}^K  \frac{1}{2\pi}\int_0^{2\pi} \rho_1^{i}(\phi+\widetilde{\phi}^i) \ d\widetilde{\phi}^i
= \frac{1}{K}\sum_{i=1}^K \bar{\rho}^i\equiv \av{\bar{\rho}^i}  \ ,
\eeq
and
\beq    \label{rho2fac}
\rho_2(\phi_1-\phi_2)=
\frac{1}{K}\sum_{i=1}^K
  \frac{1}{2\pi}\int_0^{2\pi}
\rho_1^{i}(\phi_1+\widetilde{\phi}^i) \rho_1^{i}(\phi_2+\widetilde{\phi}^i)
 \ d\widetilde{\phi}^i \  .
\eeq
Here $\rho_1^{i}(\phi)$  is given by (\ref{rhoi}) and $\bar{\rho}^i$ and $v^i_n$ are given by  (\ref{rhoi0}).
The $\widetilde{\phi}^i$ is an additional common random phase,
which arises due to the event-by-event fluctuation of the reaction plane.
Note that we also add an additional averaging over this phase for each string configuration,
which corresponds to the azimuthal rotation of a given string configuration,
and  $\av{...}$ means averaging  over string configurations.
Substituting now  (\ref{rhoi}) in (\ref{rho2fac})  we get
\beq    \label{rho2b}
\rho_2(\Delta \phi)=
\av{(\bar{\rho}^i)^2} +2\sum_{n=1}^\infty
 \av{(\bar{\rho}^i v^i_n)^2} \cos (n\,\Delta \phi) \  .
\eeq
Then
\beq \label{Ra}
C_2(\Delta \phi)
=\frac{2}{\av{\bar{\rho}^i}^2}
\sum_{n=1}^\infty \av{(\bar{\rho}^i v^i_n)^2} \cos (n\Delta \phi)+C_0
=2\sum_{n=1}^\infty \av{(\frac{\bar{\rho}^i}{\av{\bar{\rho}^i}} v^i_n)^2} \cos (n\Delta \phi)+C_0
 \ ,
\eeq
where
\beq \label{C0}
C_0=\frac{\av{(\bar{\rho}^i)^2} - \av{\bar{\rho}^i}^2 }
{\av{\bar{\rho}^i}^2}  \ .
\eeq
In section 2 we have emphasized the importance
of the observation of the common ``pedestal'' value
in $C_2$, which in that model was equal to the variance of the number
of emitters divided by the square of their mean number
(\ref{C2Lam1}).
In the present model by (\ref{C0}) we see again that the value
of this constant $C_0$ is equal to the variance
of the mean (for given string configuration $i$) multiplicity $\bar{\rho}^i$  (\ref{rhoi0})
from one string configuration to another divided by the square
of the averaged multiplicity.

We see also that the ridge like structure in (\ref{Ra}) is expressed
through the same Fourier harmonics $a_n^i$ and $b_n^i$ (\ref{ain}),
as the azimuthal flows $v_n$ (\ref{vn}),
and  the mean multiplicity $\bar{\rho}^i$ for given string configuration $i$ (\ref{rhoi0}) .

%%%%%%%%%%%%%%%%%%%%%%%%%%%%%%%%%%%%%%%

Further rough evaluation of (\ref{Ra}) is possible only if we will consider
that the mean multiplicity $\bar{\rho}^i$ weakly depends on string configuration $i$:
$\bar{\rho}^i \approx \av{\bar{\rho}^i} =const$, which is poorly justified assumption.
Under this assumption $C_0=0$ and
\beq
\label{Ra1}
C_2(\Delta \phi)= 2\sum_{n=1}^\infty \av{(v^i_n)^2} \cos (n\Delta \phi)= 2\sum_{n=1}^\infty (v^{ms}_n)^2 \cos (n\Delta \phi)
 \ .
\eeq
Note that even in this very rude approximation the $C_2(\Delta \phi)$
is expressed not directly through the flows (\ref{vn}), but  through the "mean squared flows"  $v^{ms}_n$:
\beq
\label{vnms}
v^{ms}_n\equiv\sqrt{\av{(v^i_n)^2}}=\sqrt{\frac{1}{K}\sum_{i=1}^K  (v^i_n)^2}
=\sqrt{\frac{1}{K}\sum_{i=1}^K  (a^{i2}_n +b^{i2}_n)}
\eeq

By (\ref{Ra}) and (\ref{Ra1}) we see that in any model the
nonzero direct flow $v_1$
%in the framework of a simple model (without initial internal correlations)
%that the final state interactions (FSI) can lead through the direct flow to the formation of
leads to
the forward ridge structure in the resulting two-particle correlation function.
%%%%%%%%%%%%%%%%%%%%%
%

%%%%%%%%%%%%%%%%%%%%%%%%%%%%%%%%%%%%%%%
%%%%%%%%%%%%%%%%%%%%%%%%%%%%%%%%%%%%%%%

\end{document}